----------
X-Sun-Data-Type: default
X-Sun-Data-Description: default
X-Sun-Data-Name: Hong.tex
X-Sun-Content-Lines: 687

\documentstyle[aps,preprint]{revtex}
\begin{document}
\title{\bf Hydrodynamic Description of Granular Convection}
\author{Hisao Hayakawa$^{1}$\thanks{E-mail address: hisao@engels.physics.uiuc.edu}
, Su Yue$^2$ and Daniel C. Hong$^2$\thanks{
E-mail address: dh09@lehigh.edu}}
\address{$^{1}$Department of Physics, University of Illinois at Urbana-Champaign,
  1110 West Green Street, Urbana, IL61801-3080 \\
$^{2}$ Department of Physics, Lehigh University, Bethlehem, Pennsylvania 18015
}
\maketitle
\begin{abstract}
We present a hydrodynamic model that captures the essence of granular
dynamics in a vibrating bed.  We carry out the linear stability analysis and
uncover the instability mechanism that leads to the appearance of
the convective rolls via a supercritical bifurcation of a bouncing solution.  
We also explicitly determine the
onset of convection as a function of control parameters and confirm our
picture by numerical simulations of the continuum equations.
\noindent
PACS numbers: 47.20.-k, 46.10+z, 81.35.+k
\end{abstract}
\vskip 0.3 true cm
Granular materials in a container subjected to vertical vibrations
display interesting nonlinear dynamical behaviors.\cite{1,Gidaspow,2,3,4,5} 
Nothing really happens for $\Gamma = A\omega^2/g < 1$, where A and $\omega$ 
are the frequency and the amplitude of the oscillations and g is the
gravitational constant.
  But for 
$ 1<\Gamma $,
the granular materials collectively move up and down, which we term the
uniform bouncing\cite{6}, 
until $\Gamma$ reaches the critical value $\Gamma_c$ beyond which
such a uniform bouncing motion
becomes unstable and the permanent convective rolls develop
inside the bulk.\cite{2,3,4}
Recent studies have revealed
further complexity of this problem for values of $\Gamma$ much larger than
one, where reverse convection\cite{4}
and bubble formation\cite{5} have been observed.   
Current efforts to understand the experiments of granular dynamics
\cite{2,4,5}
have mostly focused on
large scale Molecular Dynamics (MD) simulations.\cite{3}
While successful in reproducing
convection cells and some of the experimental results,
such studies have limitations in understanding the analytic
structure of the instability mechanism and/or its subsequent dynamic evolution.
There have been a handful of
attempts in the past to derive continuum equations for granular dynamics
notably by Jenkins and Savage for rapid
granular flow problems \cite{7} and by Haff for
vibrating beds \cite{8}, but these studies have been mostly
confined either to simple cases of one dimensional oscillations in an infinite
system where pressure inside the grains behave like a fluid\cite{8}, or to cases
where explicit assumption has been made regarding the
Gaussian nature in velocity distribution of grains \cite{7},
which have been shown to
break down in a dense granular system \cite{9}.  There also has been
a recent attempt by
Bourzutchky and Miller\cite{10} who have utilized the
Navier Stoke equation along
the similar line of Haff \cite{8}
and have reproduced
numerically the convective rolls.  However, we find it difficult
to imagine that the hydrodynamic pressure term ($\rho gz$)
exists to cancel the gravity term
inside the granular materials that undergo vertical
vibrations.
\vskip 0.2 true cm
The purpose of this Letter is two fold: we first propose a dynamic model that
is simple enough to make progress in analytic studies, yet
captures, in our opinion, the essence of granular dynamics of vibrating beds.
Second, we demonstrate that the correct way 
of studying the convective instability
is to carry out the stability analysis
around the {\it bouncing solution} and
explicitly determine the onset of convection as a function of external
parameters.
We will also
present numerical results to confirm our predictions.
\vskip 0.3 true cm
\noindent
{\bf Equations of motion}: Our starting point is the recognition that the
most fundamental aspect of the vibrating bed,
apart from the obvious fixed bed solution with no external driving, is
the existence of a uniform bouncing of a collection of particles.  
Such a bouncing 
solution can be either a solid block 
inside the bed or
a fluidized state with a slightly expanded volume yet with no
internal degrees of freedom. This assumption 
is consistent with 
observations in MD \cite{3} where surface fluidization rapidly spreads out into
bulk regions when surface fluidization is suppressed.  
In such a case, the bouncing solution can be 
represented by a motion of a ball on a vibrating platform.  For small
$\Gamma$,
no exotic motion such as chaotic motion
is expected to occur for such a ball \cite{11}.
We further assume that the restitution constant of the collection of particles
is zero to represent the relaxation of inside collection of particles.
  In such a case,  the relative position of the
ball with respect to the bottom plate, $\Delta (t)$, is given by:
\begin{equation}
\label{Delta(t)}
\Delta (t) = \Gamma (\sin{ t_0} - \sin {\it t}) + \Gamma \cos{ t_0}
(t-t_0)
-{\frac{1}{2}}(t-t_0)^2 
\end{equation}
in the unit of $g=\omega=1$, where the ball starts to bounce at $t_0$ on
the bottom plate, whose position at time $t$ is given by $
\Gamma \sin {\it t}$ in the experimental frame.
The bouncing solution is then described by the relative
speed between the plate and the ball:
$ V_{rel}\equiv d\Delta(t)/dt $.
Since $\Delta (t)$ cannot be negative, the ball launched upward
on the plate at $t_0$ falls back to the plate at $t_1$(i.e,$\Delta(t_1)=0$)
 and stays there until
$t=t_0+2\pi$ from our assumption of the zero restitution constant.  
The ball is then relaunched and obeys (1) again.
For later use, we determine ($t_0=\sin^{-1}(1/\Gamma),t_1$)
for different values of $\Gamma$.  For example, $(t_0,t_1)=(1.181,2.88225)$
for $\Gamma=1.1$ and $(t_0,t_1)=(0.524,5.18)$ for $\Gamma=2.0$.
When we expand $\Delta(t)$ around $t_0$ we obtain 
$\Delta(t)\simeq (\Gamma/6) \cos(t_0)(t-t_0)^3>0$, where
$\cos(t_0)>0$ from the launching 
condition $d^2V_{rel}/dt^2>0$.
Hence,
there is no solution of $\Delta(t)=0$ around $t=t_0$ except for $t=t_0$.
Therefore, the bouncing motion starts from the finite $t_1-t_0$.
One can now readily derive 
the equation of motion for the vertical
coordinate z for the bouncing motion of a granular block:
$\ddot z = (-1 + \Gamma \sin{\it t})\theta (-1 + \Gamma \sin{\it t})$
where the $\theta(x)=1$ for $x>0$ and $\theta(x)=0$ for otherwise.

In order to describe the motion of a granular block 
in the presence of internal degrees of freedom 
such as rotation and/or translation, 
we define two coarse-grained dynamical
variables: the 
density $\rho({\bf r},t)$ and the velocity {\bf v}({\bf
r},t) of the granular system.   In a box fixed frame,
$\rho$ and {\bf v} then should satisfy
the continuity and the momentum equation:
\begin{eqnarray}
 \partial_t\rho + \nabla\cdot(\rho {\bf v})&=&0 \label{mass-cont}\\
 \partial_t{\bf v} + ({\bf v}\cdot\nabla){\bf v} & =& \hat z(\Gamma \sin{\it t}
-1-\lambda)
- {\frac{1}{\rho}}\nabla P + {\frac{1}{R}}[\nabla^2{\bf v} + \chi\nabla(
\nabla\cdot{\bf v})]
\label{momentum-cont}\end{eqnarray}
where $\hat z$ is the unit vector in the vertical 
direction and $\lambda$ is a Lagrange multiplier.  
$\lambda=0$ for free motion and $\lambda=\Gamma \sin t-1$ for stationary state.
Note that the first term in (\ref{momentum-cont})
 is due to the uniform bouncing and the
third term is the energy dissipation effectively represented by
the Reynolds number R and the bulk viscosity $\chi$.
 Now, the exact form of the
pressure $P$ in (\ref{momentum-cont}) is unknown for granular
materials.  Unlike fluid, for granular materials in a container supported by
the side walls, the pressure inside the bulk seems to saturate \cite{1,12}.
 In such a case, the only contribution to the granular
pressure would result from the hard sphere repulsion 
which might be
effectively represented by the Van der Waals equation:
\begin{equation}
\label{pressure}
 P = {\frac{T\rho}{1-b\rho}}
\end{equation}
where $T \approx <{\bf v}^2> $
is the granular temperature \cite{8} and $b$  is a constant of order unity.
Note that eqs.(\ref{mass-cont}) and (\ref{momentum-cont}) are
precisely the {\it compressible} Navier-Stokes equation with
two  modifications: first the hydrodynamic
pressure term is absent, which is replaced by the Van der
Waals form (\ref{pressure}), and second, the gravity term thus survives 
in the vibrating
bed and has been effectively modified to $g - A\omega^2 \sin{\it t}$
in the physical unit.  
Notice that the term $\Gamma \sin t$ appears since we have used
 the box-fixed frame. 
We now analyze eqs.(\ref{mass-cont}) and (\ref{momentum-cont}).
\vskip 0.3 true cm

\noindent {\bf Linear stability analysis}:
(a) Fixed bed solution: Fixed bed is a container
filled with grains with no external driving.
In this case, the contact force balances
out the gravity and the net force acting on each grain is zero.  So, we use
$\lambda=\Gamma \sin{\it t} -1$ in (\ref{momentum-cont}).
 In this case, 
the
solution with constant density $\rho=\rho_0$ and zero speed ${\bf v}=0$ is
stable.
\vskip 0.3 true cm
\noindent (b) Linear stability of a uniform bouncing solution:
In order to discuss the stability of the
uniform bouncing solution, $\rho=\rho_0$ and
${\bf v} = (0,0,V_{rel}(t))$, against fluctuations, we set
$\rho=\rho_0 + \rho_L$, and decompose the velocity into the vertical
and horizontal components,${\bf v}_L = ({\bf v}_{\perp,L}, w_L)$ with
${\bf v}_{\perp} = {\bf v}_{\perp,L}$ and $w = V_{rel}(t) + w_L$.
We then substitute these into dynamic equations
(\ref{mass-cont}) and (\ref{momentum-cont})
 and introduce a new coordinate to simplify the problem,
$\xi = z -\int^t V_{rel}(t')dt'$.  Upon linearization, we
obtain the following equation for the perturbed density $\rho_L$:
\begin{equation}
\label{linear}
 [\partial_t^2 - {\frac{1}{\hat R}}\nabla_{\xi}^2\partial_t - T_e
\nabla_{\xi}^2]\rho_L = 0 
\end{equation}
where $\hat R = R/(1+\chi)$.
We now solve (\ref{linear}) in 2 dimension
under the no current boundary condition at the plate and
at $z=\infty$, namely:
$\rho_L=0 \quad {\rm at} \quad x=0,L
, \quad {\rm and}
\quad z=0,\infty $
where $L$ is the dimensionless size of the box.
To satisfy these boundary conditions, we set:

\begin{equation}
\rho_L(x,y,z,t)=\rho_{L,q,m}(t)\sin[\hat \pi m x]\sin[q(\xi-S(t))]
\label{rho:mode}
\end{equation}
where $\hat \pi=\pi/L$,
$m$ is an integer,
and $S(t)=-\Delta(t)$.
 We notice that the spectrum is discrete for
$x$ direction but continuous along z direction.
We now substitute (\ref{rho:mode}) into (\ref{linear})
 and utilize the fact, $t=\tau + t_0$ with
$t_0 = \sin^{-1}1/\Gamma$.  After some algebra, we obtain the following
second order ordinary
differential equation for the amplitude $\rho_q(t)=\rho_{L,q,m}(t)$:
\begin{equation}
\label{ODE}
\ddot \rho_q +B(q) \dot \rho_q +i C(q) \dot \rho_q +D(q) \rho_q+ i E(q) \rho_q
= iL_q(\tau) \dot \rho_q+M_q(\tau) \rho_q+ i N_q(\tau) \rho_q
\end{equation}
where
\begin{eqnarray}
B(q) &= &\hat R^{-1}(\hat \pi^2 m^2+q^2) ,\quad
C(q) = 2q\sqrt{\Gamma^2-1}\\
D(q) & =& T_e(q^2+\hat\pi^2m^2)-\frac{3}{2}q^2\Gamma^2+q^2,\quad 
E(q)  =-q +\sqrt{\Gamma^2-1}\hat R^{-1} q(\hat \pi^2 m^2 +q^2)
\end{eqnarray}
and the time dependent inhomogeneous terms are:
$ L_q(\tau) = 2q[\tau+\sqrt{\Gamma^2-1}\cos\tau-\sin\tau] $,
$M_q(\tau) =  -2q^2(\Gamma^2-1)\cos\tau+ 2q^2\sqrt{\Gamma^2-1}\sin\tau
 +\frac{q^2(\Gamma^2-2)}{2}\cos(2\tau)- q^2\sqrt{\Gamma^2-1}\sin(2\tau) 
 -2q^2\sqrt{\Gamma^2-1}\tau(1-\cos\tau) -2q^2\tau\sin\tau+q^2\tau^2 $
and
$N_q(\tau)=
-q\sqrt{\Gamma^2-1}\sin\tau - q\cos\tau + \hat R^{-1}q(q^2+\hat \pi^2m^2)
(\tau+\sqrt{\Gamma^2-1}\cos\tau-\sin\tau)$.

Note that
the equation (\ref{ODE}) is valid only
between $\tau=2 n \pi $ and $\tau=\tau_0 +2 n \pi$ with $\tau_0\equiv t_1-t_0$,
 during which
grains are launched from the plate by vibrations and then undergo free fall.
Except for this region, it is easy to show $S(t)=0$ and $C(q)=E(q)=L_q(\tau)
=M_q(\tau)= N_q(\tau)= 0$ with $D(q)\to D_0(q)=T_e(q^2+\hat \pi^2 m^2)$.

The rest of the paper is devoted to discuss the condition for 
the linear stability of
(\ref{ODE}) and
numerically test the validity of such approximations.
We may be able to obtain an explicit solution of Eq.(\ref{ODE}) with the aid of
assumption that the most unstable mode is only a relevant mode.
 The condition for
instability from this treatment, however, is complicated and time dependent.
In addition, this condition is not adequate for our purpose, 
since we are interested in the behavior in time longer than
one vibrating oscillation. 
Therefore,  we may replace $L_q(\tau)$ (and for $M_q(\tau)$ and
$N_q(\tau)$ as well) by
its average value over a {\it flying time} $<L_q(\tau)>$, namely
$L_q(\tau) \simeq <L_q>=\frac{1}{\tau_0}\int_0^{\tau_0}d\tau L_q(\tau)$
and so on for $M_q(\tau)$ and $N_q(\tau)$.
Eq.(\ref{ODE}) is then reduced to a second order ordinary differential equation
 with constant coefficients.
Assuming $\rho_q\sim e^{\sigma t}$, it becomes easy to obtain the eigenvalues
$\sigma$ for the flying motion as
\begin{equation}
\label{eigenvalue}
\sigma_{\pm}=-\frac{B+i\tilde C}{2}\pm \frac{1}{2}\sqrt{(B+i \tilde C)^2-
4(\tilde D+ i \tilde E)}
\end{equation}
where $\tilde C=C(q)-<L_q>$, $\tilde D=D(q)-<M_q>$ and 
$\tilde E=E(q)-<N_q>$. The relevant branch is only $\sigma_+$.
On the other hand, the eigenvalues are reduced to 
$\sigma_{\pm}=-B/2\pm\sqrt{B^2-4D_0}/2$ for stationary states.

The averaged 
instability condition over one oscillation cycle is then 
the average of $Re[\sigma]>0$.
For this purpose, we introduce a function:
\begin{equation}
\label{instability}
\sigma_{eff}(q)=\tau_0\{(\tilde E-\frac{B \tilde C}{2})^2-
B^2(\tilde D+\frac{\tilde C^2}{4})\} +(2\pi-\tau_0)(-B^2 D_0) .
\end{equation}
where the first term 
is the instability condition
for (\ref{eigenvalue}) 
multiplied by the time period,$\tau_0$, in which particles can move
freely\cite{HKT}, while the second term is that with $S(t)=0$.
If the function 
$\sigma_{eff}(q)>0$ for any $q$, it signals the instability of the
uniform bouncing solution.
For finite system, $\sigma_{eff}(0)=-2\pi^7T_e/(\hat R^2L^6)<0$.
Thus, the convection will
disappear for infinite systems, which agrees 
with MD simulations\cite{1,2,Taguchi}.  Equivalently, convection also 
disappears in the limit of large $R$, i.e. either the particles are too smooth
or the kinetic energy is too small
to provide the necessary driving force among grains.
The set of parameters that corresponds to physical situations might be
:$\hat R\sim 2, 
T_e\sim 3$ and $L=10$, because (i)
the linear size of the box
$L_r=L g/\omega^2\simeq 0.6 [cm]$ for $\omega\simeq 20 [Hz]$, 
(ii) $T\sim \tau_0^{-1}\int_0^{\tau_0}V_{rel}^2(\tau)d\tau \sim 3$, 
(iii) the kinetic viscosity for granular fluid is evaluated by
$\nu_s\simeq 5\times 10^{-3} [m^2/s]$\cite{Gidaspow} and
the definition of $R=U L_r/\nu_s\sim 2$ 
with the aid of
the characteristic velocity $U\sim \sqrt{V_{rel}^2} g/\omega\sim 10 cm/sec$
 in the physical unit.  But for pure numerical reasons, we choose
$\hat R = T_e=10$ and $L=10$.
For this set of parameters, we first solve $\Delta(t)=0$ numerically to
determine $t_1$, and then compute $\sigma_{eff}(q)$ as a function of q.
As demonstrated in Fig.1,
$\sigma_{eff}(q)$ is convex and thus
has a maximum, $\sigma_m$, at a particular value of
$q$.  For  
$\Gamma\simeq 1$, $\sigma_m <0$ and thus 
$\sigma_{eff}(q)<0 $ for all $q$ and the
bouncing solution is stable.  But as we increase $\Gamma$ further to the
 critical value, $\Gamma_c$, $\sigma_m$ moves upward crossing
 zero and becomes positive,
in which case $\sigma_{eff}(q)>0$ for a band of $q$.  In this case, the bouncing
solution becomes unstable and we expect the convective rolls to appear.
The onset of convection is then determined by setting, $\sigma_m(\Gamma_c)=0$.
 For $L=10$,$R=T_e=10$, we find $\Gamma_c\simeq 1.12$ and the selcted wave number 
is about $q_c= 0.22$. The most unstable wave number $q_m$ gradually shifts
with the increase of $\Gamma$.
We now check the validity of our picture by numerical simulations.


\vskip 0.3 true cm
{\bf Numerical Results}:
We have solved (2)-(4) numerically 
in two dimension with
no slip boundary conditions at the side walls as well as at the top and the 
bottom plates. Note that the top plate suppress complicated surface motion
of vibrating beds and allow us to use the simplified picture.
Since the granular fluid is confined in a box, we
do not introduce $\lambda$ explicitly in the simulations.  As a result, $S(t)
\approx 0$ after a grain lands on a plate in the average bouncing state.  The
absence of $\lambda$ and the presence of the top wall is expected to cause the 
appearance of the bouncing solution for $\Gamma \le 1$ 
in contrast with the real
situation.  But since
the
linearized eq.(\ref{ODE}) with $S(t) = 0$ is identical to that with non-zero
$\lambda$, omitting $\lambda$ would not change the essence of the
dynamics.  In the same spirit, we have ignored $\chi$ and $b$ in our 
simulations.
Our simulation results are presented in Fig.2 for two different values of
$\Gamma$, $\Gamma<\Gamma_c$ and $\Gamma>\Gamma_c$.
In the former case,
the bouncing solution is expected to appear inside the bed and  
the density and the velocity at
a given point oscillates with the same frequency of the vibration.(Fig.2a)
Upon increasing $\Gamma$ further to $\Gamma = 1.2$, which is
beyond the predicted $\Gamma_c = 1.12$ determined by 
(11), we find that
 the bouncing solution has disappeared 
and the permanent convective rolls have developed inside the bulk (Fig.2b).
The wavelength of the most unstable mode by the linear stability analysis 
is about $q_m\approx 0.4$, which is
not far from the actual wavelength of the convective rolls: $q=2\pi/\lambda = 
2\pi/L\approx 0.6$.  

In passing, we briefly mentioned the difference between the granular
beds and the water beds.  The latter is shown to exhibit the Faraday 
instability on the air-water interface.\cite{Faraday}  The crucial difference between these
two systems lie in the pressure term: for the water bed, since the
water is an incompressible fluid, the hydrodynamic
pressure term $\rho g z$ precisely cancels the gravity term in the fluid
equation, thus suppressing the motion inside the bulk, while the absence
of the hydrodynamic pressure term produces the convective
instability in the bulk for the
granular beds.   We will present the details of our analysis including
the weakly nonlinear analysis elsewhere, which will highlight
the differences between the two.


\vspace*{0.5cm}
{\bf Acknowledgment}
The authors wish to thank Y-h. Taguchi and D.A. Kurtze for stimulating discussion.
This work is, in part, supported by the Foundation for 
Promotion of Industrial Science in Japan
and the U.S. National Science Foundation 
through grant number NSF-DMR-93-14938. 

\vfill\eject

\vfill\eject
\noindent {\bf Figure Captions}
\vskip 0.5 true cm
\noindent Fig.1. The effective growth rate $\sigma_{eff}(q)$ as a function of
the wave number q for $\Gamma =1.05$(diamond),for which
$\sigma_{eff}(q)<0$ for all values of q, while
for $\Gamma=1.2>\Gamma_c = 1.12$, 
$\sigma_{eff}(q)$ becomes positive for a band of q(square).  
$\Gamma_c$ is determined by the
condition that the maximum of $\sigma_{eff}(q)$ becomes zero
 at $\Gamma_c$.(cross)
The parameters used are: $T_e=R=10$ and $L=10$.
\vskip 0.3 true cm
\noindent Fig.2 (a) A bouncing solution.
The speed $v_z$ at a given point is plotted as a function 
time for $\Gamma = 0.9$.
 (b) For $\Gamma =1.2>\Gamma_c=1.12$,
the bouncing solution becomes unstable and
the permanent convective rolls appear inside the box.  The arrows are the
velocity vectors pointing upward.
The parameters used in simulations for (a) and
(b) are the same as those in Fig.1.





\begin{references}
\bibitem{6} Uniform bouncing is a periodic
motion of a granular block that might
be a solid or a fluid with slightly expanded volume. 
While surface motions of particles are more complicated than 
what we have explained here, if we suppress surface motions by the top plate,
the simulation by Taguchi\cite{3} is consistent with our picture.
\end{references}
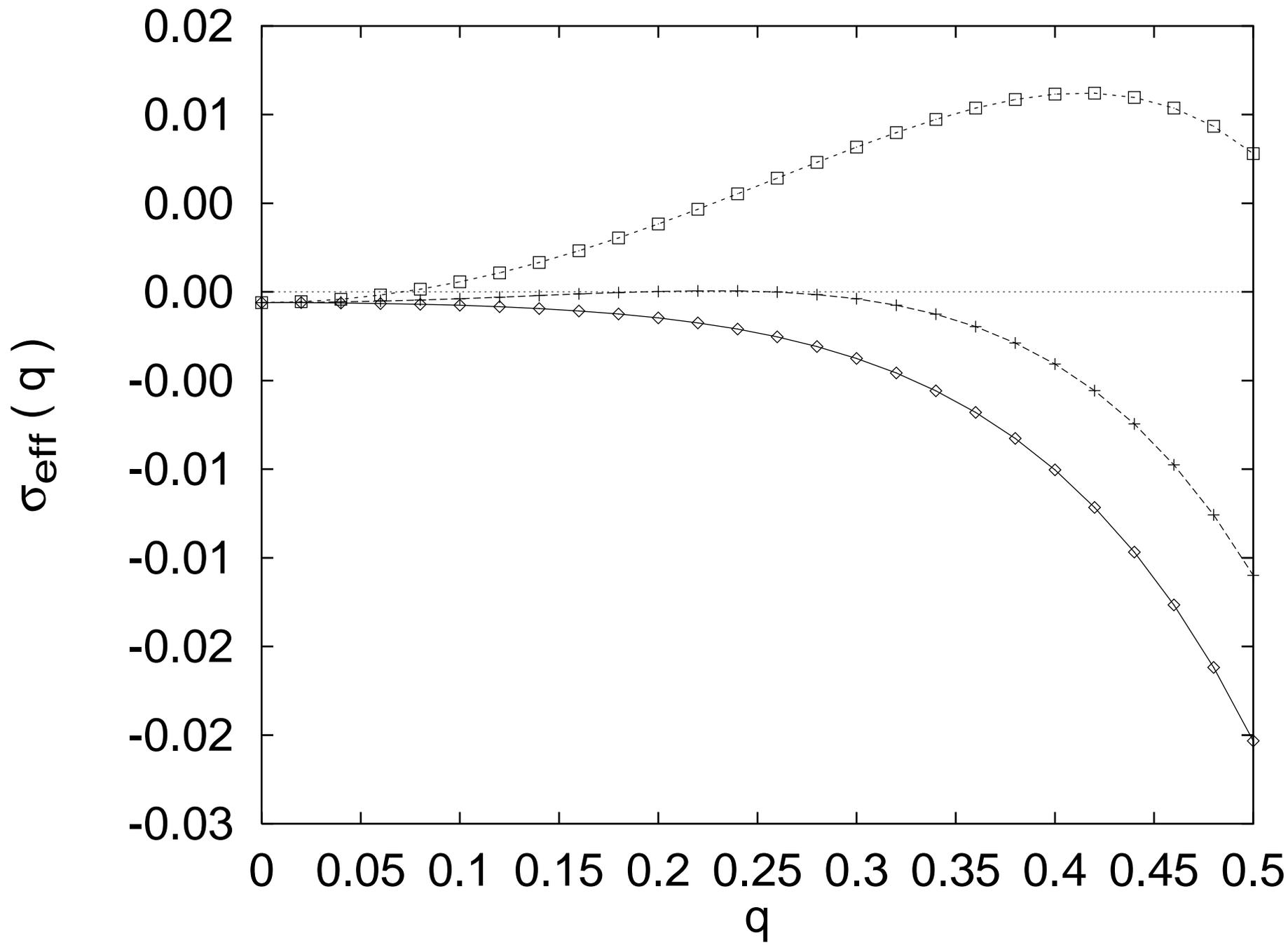
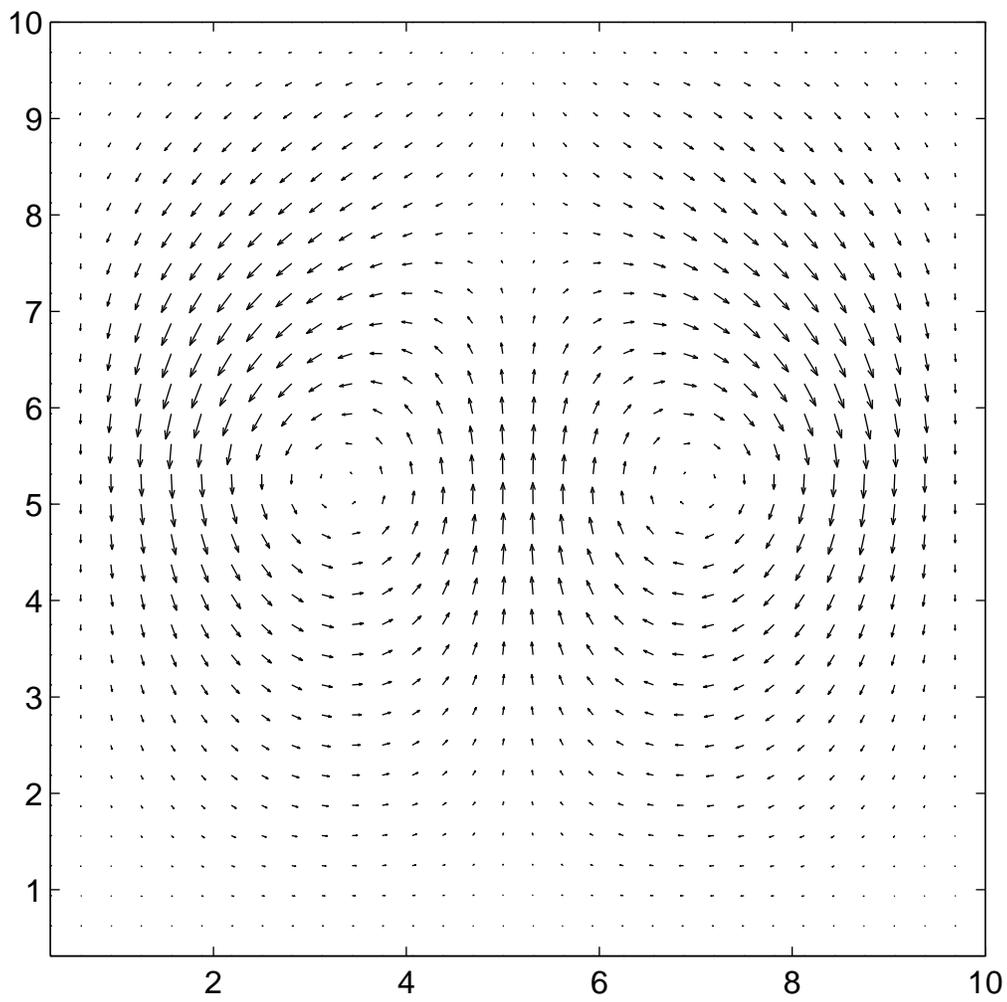
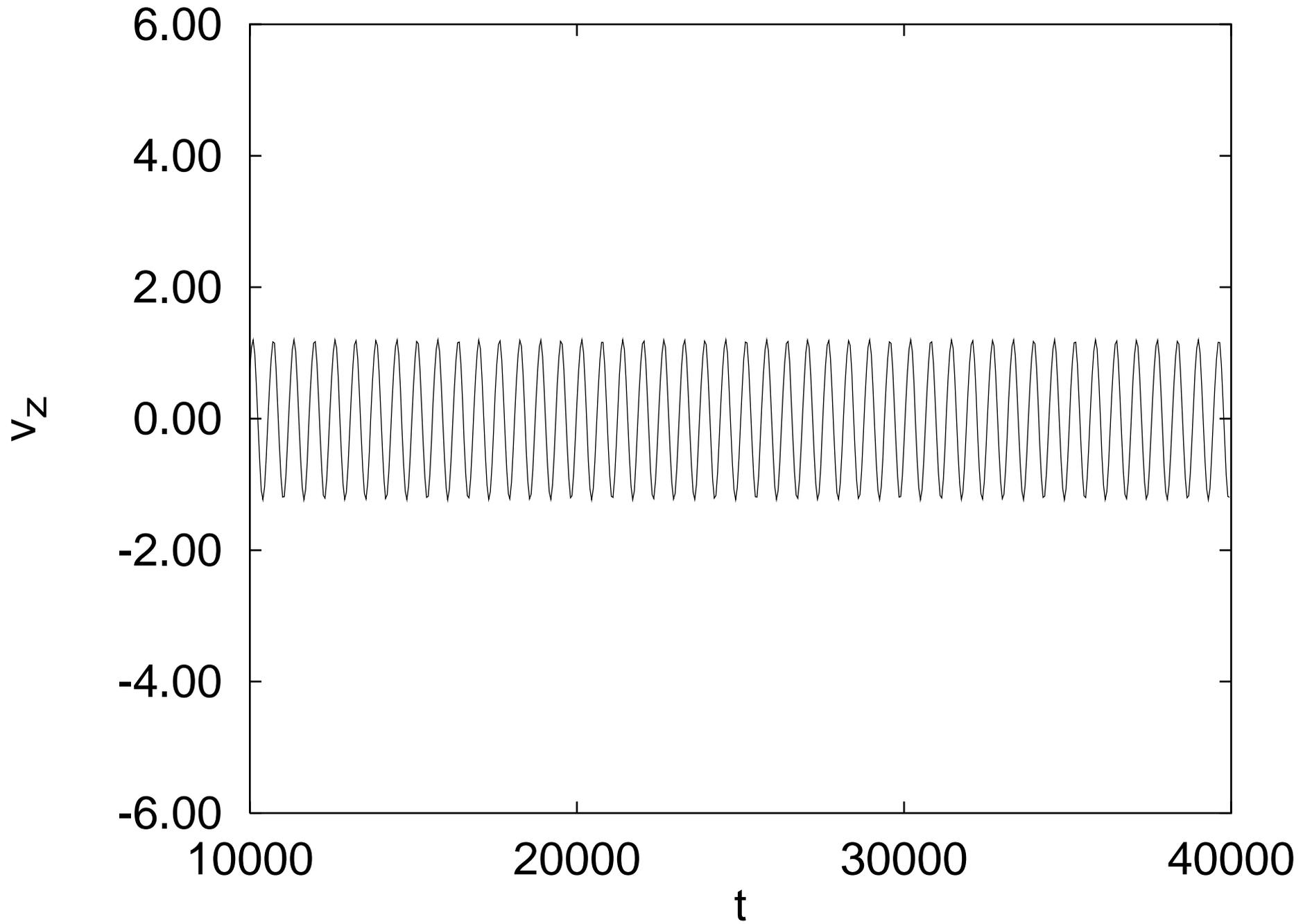
\end{document}